\documentclass[aps,pra,reprint,superscriptaddress,floatfix]{revtex4-1}
\usepackage{amsmath}
\usepackage{amssymb}
\usepackage{graphicx}
\usepackage{color}
\newcommand{\bea}{\begin{eqnarray}}
\newcommand{\eea}{\end{eqnarray}}
\newcommand{\bse}{\begin{subequations}}
\newcommand{\ese}{\end{subequations}}

\begin{document}

\title{Effect of controlled point-like disorder induced by 2.5 MeV electron irradiation on nematic resistivity anisotropy of hole-doped (Ba,K)Fe$_2$As$_2$}

\author{M.~A.~ Tanatar}
\affiliation{Ames Laboratory, USDOE, Ames, Iowa 50011, USA}
\affiliation{Department of Physics and Astronomy, Iowa State University, Ames, Iowa 50011, USA}
\email{tanatar@ameslab.gov}

\author{Erik I.~Timmons}
\affiliation{Ames Laboratory, USDOE, Ames, Iowa 50011, USA}
\affiliation{Department of Physics and Astronomy, Iowa State University, Ames, Iowa 50011, USA}
%\email{erikt@iastate.edu}

\author{M.~Ko\'nczykowski}
\affiliation{Laboratoire des Solides Irradis, CEA/DRF/lRAMIS, Ecole Polytechnique,
CNRS, Institut Polytechnique de Paris, F-91128 Palaiseau, France}
%\email{marcin.konczykowski@polytechnique.edu}

\author{O.~Cavani}
\affiliation{Laboratoire des Solides Irradis, CEA/DRF/lRAMIS, Ecole Polytechnique,
CNRS, Institut Polytechnique de Paris, F-91128 Palaiseau, France}
%\email{olivier.cavani@polytechnique.edu}

\author{Kyuil Cho}
\affiliation{Ames Laboratory, USDOE, Ames, Iowa 50011, USA}
%\email{kcho@ameslab.gov}

\author{Yong Liu}
\affiliation{Ames Laboratory, USDOE, Ames, Iowa 50011, USA}
%\email{yongliu31@outlook.com}

\author{T.~A.~Lograsso}
\affiliation{Ames Laboratory, USDOE, Ames, Iowa 50011, USA}
\affiliation{Department of Material Science and Engineering, Iowa State University, Ames, Iowa 50011, USA}
%\email{lograsso@ameslab.gov}

\author{R.~Prozorov}
\affiliation{Ames Laboratory, USDOE, Ames, Iowa 50011, USA}
\affiliation{Department of Physics and Astronomy, Iowa State University, Ames, Iowa 50011, USA}
%\email{prozorov@ameslab.gov}

\date{\today}

\begin{abstract}

In-plane anisotropy of electrical resistivity was studied in samples of the hole-doped Ba$_{1-x}$K$_x$Fe$_2$As$_2$ in the composition range $0.21 \leq x \leq 0.26$ where anisotropy changes sign. Low-temperature ($\sim$20~K) irradiation with relativistic 2.5 MeV electrons was used to control the level of disorder and residual resistivity of the samples. Modification of the stress-detwinning technique enabled measurements of the same samples before and after irradiation, leading to conclusion of anisotropic character of predominantly inelastic scattering processes. Our main finding is that the resistivity anisotropy is of the same sign irrespective of residual resistivity, and remains the same in the orthorhombic $C_2$ phase above the re-entrant tetragonal transition. Unusual $T$-linear dependence of the anisotropy $\Delta \rho \equiv \rho_a(T)-\rho_b(T)$ is found in pristine samples with $x=$0.213 and $x=$0.219, without similar signatures in either $\rho_a(T)$ or  $\rho_b(T)$. We show that this feature can be reproduced by a phenomenological model of R.~M.~Fernandes {\it et al.} Phys. Rev. Lett. {\bf 107},217002 (2011). We speculate that onset of  fluctuations of nematic order on approaching the instability towards the re-entrant tetragonal phase contributes to this unusual dependence.

\end{abstract}

\maketitle

\section{Introduction}

\begin{figure}
\includegraphics [width=3.0in]{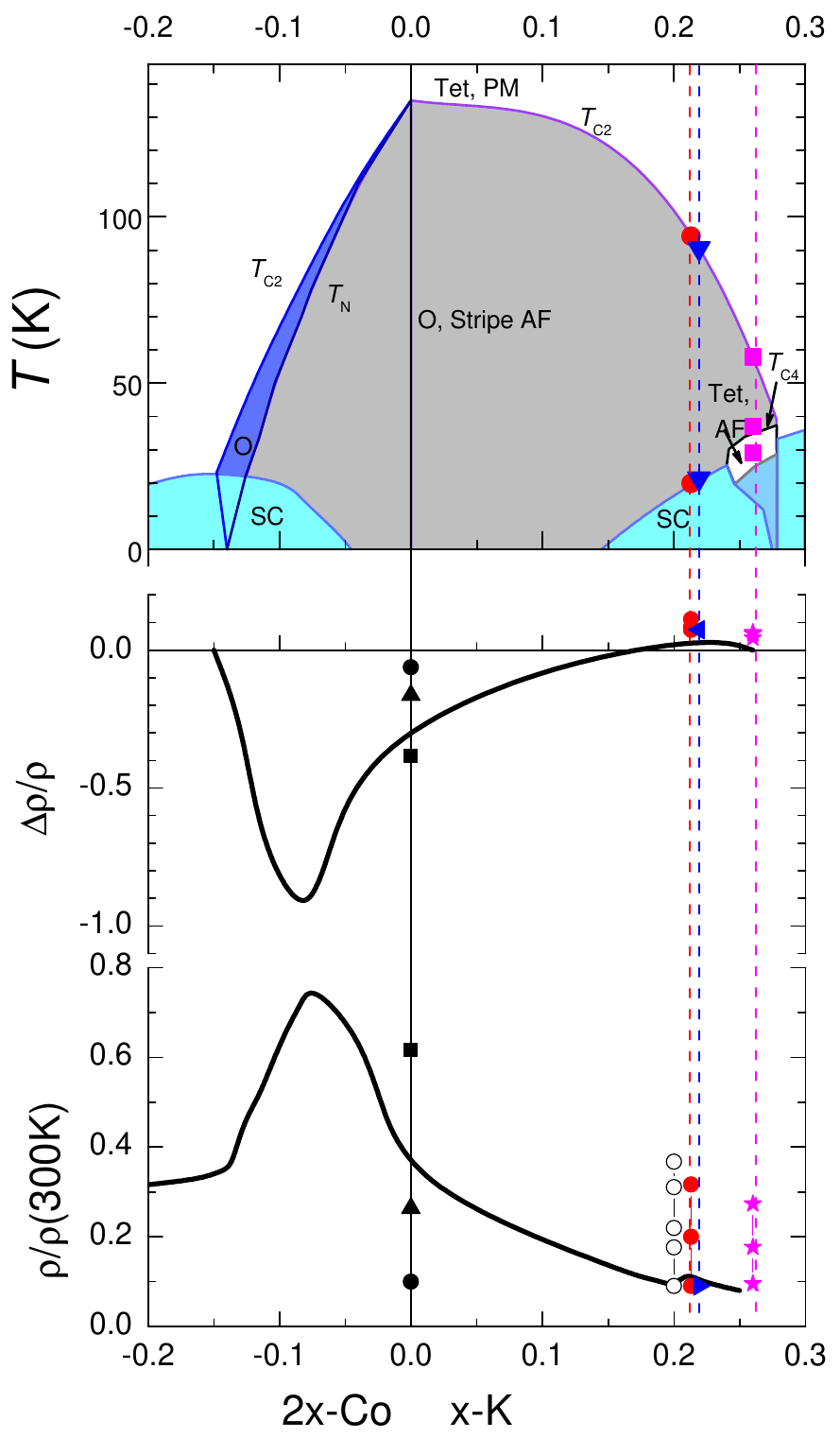}
\caption {(Color online) Top panel. Summary phase diagram of electron-, Ba(Fe$_{1-x}$Co$_x$)$_2$As$_2$,  and hole-, Ba$_{1-x}$K$_x$Fe$_2$As$_2$, doped iron based superconductors. Red, blue and magenta points are $T_{C2}$, $T_c$ and $T_{C4}$ of the pristine samples $x=$0.213, 0.219 and 0.260 respectively, used in this study. Middle panel shows composition dependence of the low-temperature resistivity anisotropy, $\Delta \rho/\rho$, where $\Delta \rho =\rho_a-\rho_b$. Black solid symbols in the middle and bottom panels show effect of residual resistivity on resistivity anisotropy at low temperatures in parent BaFe$_2$As$_2$, squares after \cite{detwinning}, triangles after  \cite{BlombergJPCM} and circles after \cite{Ishida}. Red, blue and magenta symbols are from this study. Bottom panel shows evolution of the resistivity ratio, $\rho(0)/\rho(300K)$ taken as a proxy of the residual resistivity. Open black circles are for the samples with $x=$0.20 subjected to electron irradiation \cite{npjQM}, red and magenta symbols from the samples studied in this article, $x=$0.213 and $x=$0.260, in the sign reversal composition range. Blue is for the sample with $x=$0.219 studied only in the pristine state. }
\label{phaseD}
\end{figure}

Studies of in-plane anisotropy of electrical resistivity in iron-based superconductors are performed on stress-detwinned samples \cite{detwinning,Ian1} creating preferential orientation of orhthorombic domains \cite{domains}. The resistivities for principal orthorhombic directions, $a$ and $b$, $\rho_a(T)$ and $\rho_b(T)$, and their difference $\Delta \rho \equiv \rho_a-\rho_b$ referred to as anisotropy, reveal several unusual features. The resistivity of the parent BaFe$_2$As$_2$ is lower for the long $a-$ axis, $\rho_a < \rho_b$, corresponding to the antiferromagnetic chains in the stripe magnetic structure. The anisotropy increases with electron doping [and suppression of the orthorhombic distortion $\delta= (a-b)/(a+b)$], taking maximum near optimal doping on electron-doped side \cite{Ian1}. The anisotropy changes sign on the hole-doped side \cite{BlombergNC}, with $\rho_a>\rho_b$, see phase diagram, Fig.~\ref{phaseD}. The mechanism of this sign change in the electronic transport attracts notable interest, since contributions from both elastic scattering due to impurities/defects \cite{elastic1,elastic2} and inelastic scattering on magnetic excitations \cite{inelastic1,inelastic2} and phonons can be anisotropic.

The magnitude of the anisotropy strongly depends on sample residual resistivity, as found in the study  on the annealed samples \cite{Ishida,Uchida2Ru,BlombergJPCM}. It was argued \cite{inelastic2} that the sign change of the resistivity anisotropy can be caused by dramatic difference in the levels of disorder scattering on the electron-doped side in Ba(Fe$_{1-x}TM_x$)$_2$ ($TM$= Co, Ni, Rh, Ir \cite{pseudogap,CBreview}) and the hole-doped side in Ba$_{1-x}$K$_x$Fe$_2$As$_2$ \cite{XLuo,Hassinger1,YLiucrystals,XChen}, as summarized in the bottom panel of Fig.~\ref{phaseD}. Indeed, substitution in the electronically active Fe sites introduces high level of scattering, with residual resistivity extrapolating to 100 $\mu \Omega cm$ or so close to optimal doping.
The K-substitution in Ba$_{1-x}$K$_x$Fe$_2$As$_2$ proceeds in electronically inactive Ba site and the residual resistivities are typically close to 30 $\mu \Omega cm$. This difference may imply that the sign may be the same for all the phase diagram.

Another consideration regarding the origin of the sign change is related to approaching the composition range of the re-entrant tetragonal $C_4$ phase \cite{Hassinger1,BohmerNC,Avci}. At ambient pressure for compositions $x<\sim 0.24$ the samples of Ba$_{1-x}$K$_x$Fe$_2$As$_2$ undergo simultaneous structural (tetragonal to orthorhombic) and magnetic (paramagnetic to stripe antiferromagnetic) transition below $T_{C2}$ (see phase diagram Fig.~\ref{phaseD}). For $x>0.24$ a sequence of phase transitions is observed, with re-entrance of the tetragonal phase below $T_{C4}$ with complicated antiferromagnetic structure \cite{C4magnetic}. This phase was not known at the time of resistivity anisotropy  study \cite{BlombergNC}.

We have recently succeeded controlling the residual resistivity of  the iron-based superconductors using low-temperature electron irradiation with relativistic 2.5 MeV electrons \cite{ErikPRB,npjQM,Kyuilreview} and achieving residual resistivity levels comparable to the electron-doped side, as shown in Fig.~\ref{phaseD} with open dots for $x=$0.20 \cite{npjQM}, solid red circles and magenta stars ($x=$0.213 and $x$=0.260, respectively, this study). Disorder introduced by irradiation does not change carrier density and enables disentangling effects of doping and of the substitutional disorder, which are intertwined in the electron-doped Ba(Fe$_{1-x}TM_x$)$_2$. We use this development to study electrical resistivity anisotropy in the sign change composition range $0.21 \leq x \leq 0.26$. Two compositions selected for the irradiation study were $x=$0.213 and $x=$0.260. The first sample was on the orthorhombic, $C_2$, side of the  composition boundary, the second one $x=$0.260 was in the re-entrant range. Our main finding is that the resistivity anisotropy is of the same sign irrespectively of residual resistivity, and remains the same in $C_2$ phase range above the re-entrant tetragonal transition.

\section{Experimental}

Single crystals of Ba$_{1-x}$K$_x$Fe$_2$As$_2$ were grown as described in detail in Ref.~\onlinecite{YLiucrystals}. Large, above 5$\times$5 mm$^2$ surface area crystals were cleaved on both sides to a thickness of typically 0.1 mm to minimize the variation of the K-content with thickness.  The crystals from two different batches were used in this study with average  compositions $x_{av}$=0.22 and 0.25, as determined from the electron-probe microanalysis with wavelength dispersive spectroscopy (WDS). The large slabs were cut using wire saw along the tetragonal [110] direction. Several cuts were made side by side to achieve the  closest similarity of the sample properties. Multiple samples cut were mounted for four probe resistivity measurements. Contacts to the samples were tin-soldered \cite{SUST,patent}.
These contacts are strong enough to withstand multiple irradiation measurements \cite{ErikPRB} and the applications of stress \cite{hook}. Samples were pre-characterized by the electrical resistivity measurements, to ascertain reproducible properties. Despite identical WDS composition, samples revealed some variation in positions of features in $\rho(T)$ curves at the concomitant structural/magnetic transition $T_{C2}$ and superconducting $T_c$. We account for this variation using polynomial fits of $T_{C2}(x)$ and $T_c(x)$ \cite{BaKcaxis}. This was particularly important for samples from the batch with $x_{av}=$0.25, as these show some variation of the  positions of $T_{C2}$ and $T_{C4}$ features in $\rho(T)$ even between the crystals cut from the same slab. Samples selected for irradiation in this study had $x=$0.213 and $x=$0.260 ($\pm0.001)$. One more sample was used for control purposes, $x=$0.219, all compositions determined from the $T_{C2}(x)$ formula \cite{BaKcaxis}. Use of $T_c(x)$ gave similar composition differences.

Due to high probability of formation of cracks during stress application, we prepared two samples of each composition. Only one sample of each composition eventually survived irradiation cycles without crack formation. The silver wires of potential contacts were used both for resistivity measurements and for stress application \cite{detwinning,BlombergSr}.  We used a specially designed device enabling easy sample mounting/dismounting and  controllable application of the tensile stress, shown in inset of  the left panel in Fig.~\ref{stressx0p213} below.
 Four-probe resistivity measurements were performed in a {\it Quantum Design} PPMS.

The low-temperature 2.5 MeV electron irradiation was performed at the SIRIUS Pelletron linear accelerator operated by the \textit{Laboratoire des Solides Irradi\'{e}s} (LSI) at the \textit{Ecole Polytechnique} in Palaiseau, France \cite{SIRIUS}.
The samples for resistivity measurements during and after electron irradiation were mounted on a thin mica plate in a hollow {\it Kyocera} chip, so that they could be moved between the irradiation chamber (in LSI) and the detwinning resistivity setup (in Ames laboratory) without disturbing the contacts.
The {\it Kyocera} chip was mounted inside the irradiation chamber and was cooled by a flow of liquid hydrogen to $T \approx 22$~K in order to remove excess heat produced by relativistic electrons upon collision. The flux of electrons amounted to about 2.7 $\mu$A of electric current through a 5 mm diameter diaphragm. This current was measured with the Faraday cup placed behind a hole in the sample stage, so that only transmitted electrons were counted. The irradiation rate was about $5 \times 10^{-6}$ C$/$(cm$^{2}\cdot $s) and large doses were accumulated over the course of several irradiation runs. The penetration depth of electrons in the hole-doped iron based superconductors is estimated as 1.3 mm \cite{webpage}, tin and silver used in the contacts have similar values, so that for samples of our dimensions the irradiation is homogeneous and there should be no shadow on the samples under the contacts. To stay on a safe side, though, the samples were positioned with electron beam incoming from the opposite to the contacts side of the samples. Throughout the manuscript we use ``pristine'' and ``unirradiated'' interchangeably to describe samples that were not exposed to electron irradiation.

Irradiation of a dose 1 C/cm$^2$ with 2.5 MeV results in about 0.07\% of the defects per iron site \cite{Kyuilreview}. The Frenkel pairs are created at about the same density in all sublattices.
It is well known that in metals, self-diffusion of interstitials is much higher than that of vacancies, especially
warming up above roughly 100 K or so   and that they mostly diffuse out and disappear at various “sinks”, like extended
defects (dislocations/disclinations) and surfaces \cite{Dines}. A much slower to relax population of vacancies remain in the
crystal in a quasi-equilibrium (metastable) state controlled by the highest temperature reached.
Resistivity measurements {\it in situ} at 22~K during irradiation in Ba$_{1-x}$K$_x$Fe$_2$As$_2$ with close composition $x=$0.20 \cite{npjQM} show linear increase with irradiation dose at a rate $\sim$ 50 $\mu \Omega$cm per 1 C/cm$^2$, decreasing to $\sim$ 30 $\mu \Omega$cm upon warming to room temperature due to defect annealing \cite{npjQM}.  The dose of defects created by electron irradiation is negligible compared with electron and hole densities in a good metal like Ba$_{1-x}$K$_x$Fe$_2$As$_2$, as verified experimentally by Hall effect measurements \cite{npjQM}.

\section{\label{Sec:rho} Electrical Resistivity}

\begin{figure}
\includegraphics [width=8.3cm]{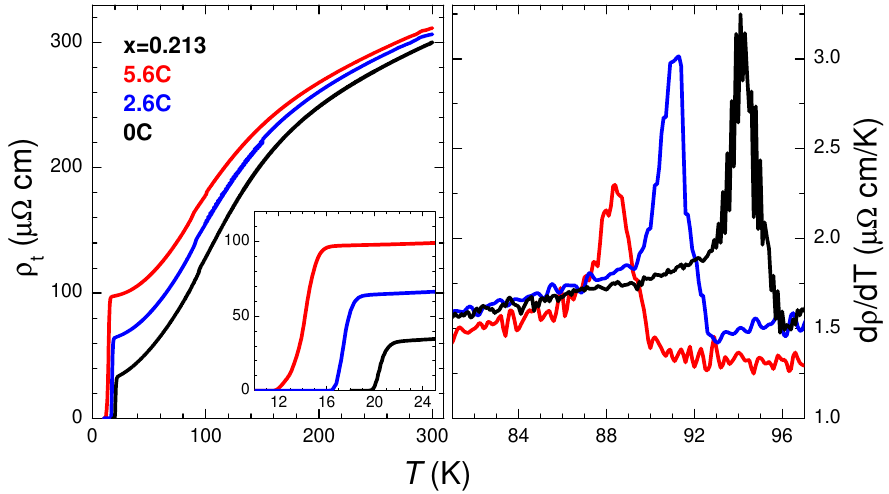}
\caption {(Color online) Temperature-dependent electrical resistivity of stress free twinned samples, $\rho_t(T)$,  of Ba$_{1-x}$K$_x$Fe$_2$As$_2$, $x=$0.213, with composition in the nematic anisotropy sign reversal range (left panel). Inset shows zoom of the superconducting transition. Black curves show data for sample before irradiation (0~C/cm$^2$), blue and red curves after irradiation with 2.6 C/cm$^2$ and 5.6 C/cm$^2$, respectively. Right panel shows temperature-dependent resistivity derivative for the data in the left panel revealing clear anomalies at the tetragonal to orthorhomic structural transition coinciding with the antiferromagnetic ordering, $T_{C2}$.  Electron irradiation monotonically increases $\rho(0)$ from $\sim$30 to $\sim$100 $\mu \Omega$cm and suppresses both $T_c$ and $T_{C2}$ at approximately the same rate. }
\label{stressfreex0p213}
\end{figure}

In Fig.~\ref{stressfreex0p213} we show evolution of the temperature-dependent resistivity of Ba$_{1-x}$K$_x$Fe$_2$As$_2$, $x=$0.213, with electron irradiation. Measurements were done in stress-free conditions in the twinned state, with resistivity denoted as $\rho_t$. The evolution is consistent with our previous studies \cite{npjQM,ErikPRB}, with suppression of the superconducting $T_c$ (inset in left panel) and of the temperature of the structural/magnetic transition, $T_{C2}$, as seen in resistivity derivative plots  (right panel). The increase of the resistivity is not constant in temperature and it is notably larger on $T \to $0, revealing notable Matthiessen rule violation. The residual resistivity increases more than by a factor of 3, from $\sim$30 to $\sim$100 $\mu \Omega$cm.

\begin{figure}
\includegraphics [width=8.3cm]{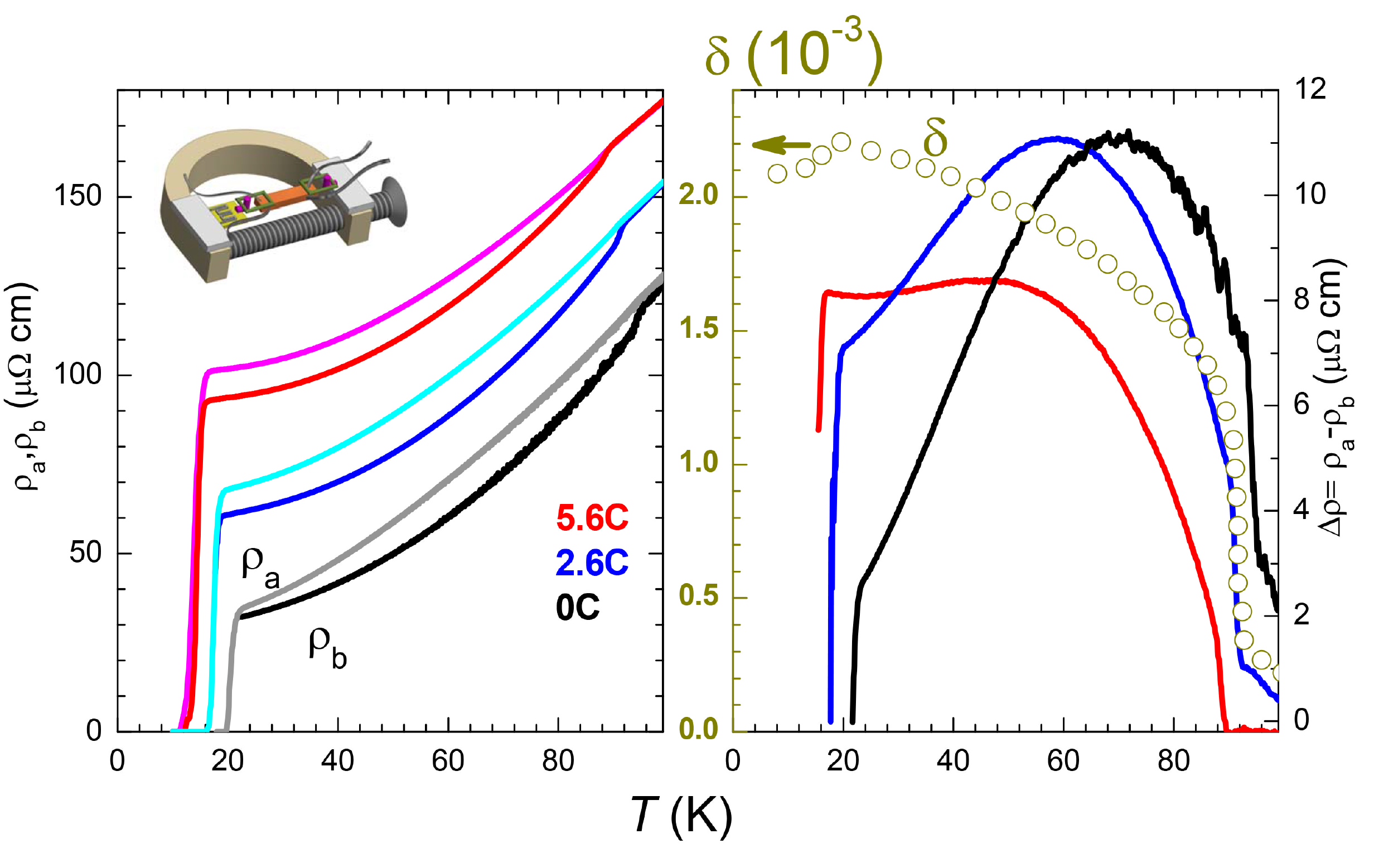}
\caption {(Color online) Temperature-dependent electrical resistivity of Ba$_{1-x}$K$_x$Fe$_2$As$_2$ sample with $x=$0.213. Two sets of curves for
each irradiation dose
represent resistivity along $a-$, $\rho_a$ (gray, cyan and magenta, top curves in the pair), and $b-$, $\rho_b$ (black, blue and red, bottom curves in the pair), directions in the conducting plane. Right panel shows temperature-dependent in-plane resistivity anisotropy, $\Delta \rho \equiv \rho_a-\rho_b$, and its evolution with irradiation. Inset in the left panel shows hook device used for detwinning experiments with multiple mounting/dismounting cycles \cite{hook}. Sample is irradiated with 2.5 MeV electrons to introduce disorder in a controlled way between stress application runs. Open dark yeallow circles show temperature dependence of the nematic order parameter, $\delta=(a-b)/(a+b)$, left axis in the right panel, in sample with $x=$0.22 in thermal expansion measurements \cite{BohmerNC}.
 }
\label{stressx0p213}
\end{figure}

On application of tensile stress using hook horseshoe device \cite{hook} sample goes into the detwinned state with predominant orientation of domains with the orthorhombic $a$-axis along the stress direction. The resistivity  increases with stress and saturates once detwinning action of stress is complete. The resistivity in this state, $\rho_a$, is shown in Fig.~\ref{stressx0p213} with grey, cyan and magenta lines for 0, 2.6 and 5.6~C/cm$^2$ samples. The bottom curves show resistivity along $b$ direction in the plane (black, blue and red curves for 0, 2.6 and 5.6~C/cm$^2$ respectively). Resistivity along $b$ direction was determined assuming equal population of domains in the stress-free sample, $\rho_t=(\rho_a+\rho_b)/2$, and $\rho_b=2\rho_t-\rho_a$.

 The in-plane resistivity anisotropy, $\Delta \rho \equiv \rho_a-\rho_b$ is shown in the right panel of Fig.~\ref{stressx0p213}. The anisotropy sign remains the same for all irradiation doses with $\rho_a>\rho_b$. The anisotropy in pristine sample (black curve in the right panel of Fig.~\ref{stressx0p213}) reaches broad maximum at about $\sim$70~K and then decreases approximately linearly down to the superconducting transition. With 2.6~C/cm$^2$ irradiation, an increase of the residual resistivity from $\sim$ 30 to $\sim$60 $\mu \Omega$cm and shift of $T_{C2}$ from 94 to 91~K, the maximum in $\Delta \rho(T)$ shifts to $\sim$60~K and some curvature starts to develop above $T_c$. The anisotropy above $T_c$ notably increases compared to the pristine sample, from $\sim$2 to $\sim$7 $\mu \Omega$cm. Finally, with 5.6~C/cm$^2$ irradiation, increase of the residual resistivity to $\sim$100 $\mu \Omega$cm and $T_{C2}$ suppression to 88~K, the maximum transforms into a plateau, starting somewhat below 60~K and continuing down to $T_c$.  This $\Delta \rho(T)$ for 5.6 C/cm$^2$ irradiated sample resembles temperature evolution of the nematic order parameter $\delta=(a-b)/(a+b)$, shown with dots (left scale in the right panel) from thermal expansion data of B\"ohmer {\it et al.} \cite{BohmerNC} for close $x=$0.22.

\begin{figure}
\includegraphics [width=8.3cm]{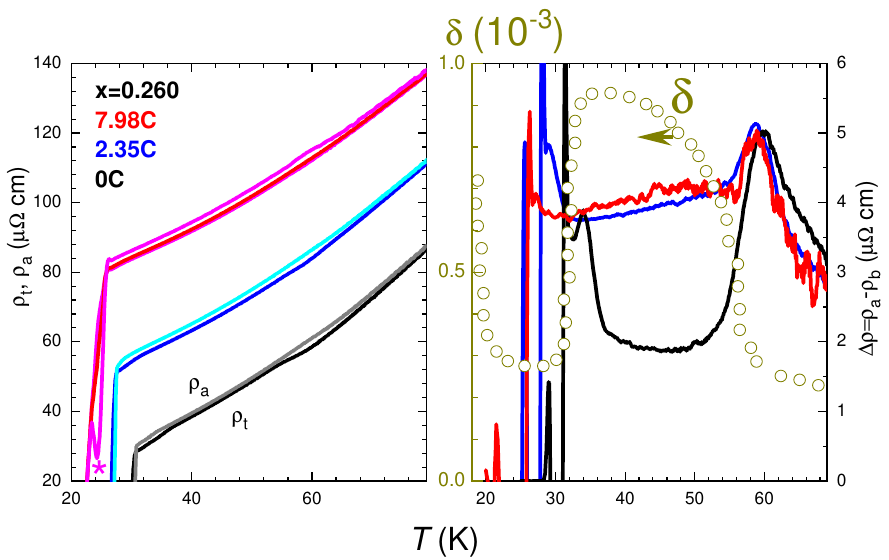}
\caption {(Color online) Temperature-dependent electrical resistivity of Ba$_{1-x}$K$_x$Fe$_2$As$_2$ sample with $x=$0.260. Two sets of curves for
each irradiation dose
represent resistivity in stress free twinned state, $\rho_t(T)$, (black, blue and red for 0, 2.35 and 7.98 C/cm$^2$ respectively) and detwinned
by application of tensile stress, $\rho_a(T)$ (gray, cyan and magenta for 0, 2.35 and 7.98 C/cm$^2$, respectively). Star marks partial cracking of the 7.98 C/cm$^2$ sample leading to a stress release.
Right panel shows temperature-dependent in-plane resistivity anisotropy, $\Delta \rho \equiv \rho_a-\rho_b$, and its evolution with irradiation. For reference we show temperature evolution of nematic order parameter, $\delta =(a-b)/(a+b)$, (open dark yellow circles, left axis in the right panel), measured with thermal expansion technique in the sample with $x=$0.262 \cite{BohmerNC}. }
\label{stress0p26}
\end{figure}

In the left panel of Fig.~\ref{stress0p26} we show evolution of the temperature dependent resistivity in  Ba$_{1-x}$K$_x$Fe$_2$As$_2$ sample with $x=$0.260. Measurements in stress-free conditions (black curve for pristine sample, blue and red for samples after irradiation with 2.35 and 7.98 C/cm$^2$, respectively) show monotonic increase of the resistivity. Note a feature at $\sim$30~K in the $\rho(T)$ curve for the sample with 7.98 C/cm$^2$ under stress (magenta line in Fig.~\ref{stress0p26}) marked with the star. Here the sample partially cracked on cooling, with the stress release. Since this crack happened after the resistivity data were taken, we were able to determine the  resistivity anisotropy as shown in the right panel 0f Fig.~\ref{stress0p26}. However in the analysis below we use the data for 2.35 C/cm$^2$ sample. The features at $T_{C2}$ (small increase on cooling below 60~K) and $T_{C4}$ (small resistivity decrease below 35~K) are very sensitive to stress, which leads to sharp anomalies in the anisotropy plot in the right panel. With irradiation the $T_{C4}$ is suppressed to at least below onset of the superconducting transition while the feature at $T_{C2}$ is nearly unaffected.

Evolution of the in-plane resistivity anisotropy in the sample $x=$0.260 is quite remarkable. The stress-induced anisotropy in the tetragonal phases above $T_{C2}$ and below $T_{C4}$ is notably larger than in the orthorhombic phase. The overall magnitude of the anisotropy is about 2 times smaller than in $x=$0.213 sample. The temperature dependence of anisotropy has little resemblance to that in $x=$0.213, with anisotropy remaining nearly temperature-independent.

\begin{figure}
\includegraphics [width=8.3cm]{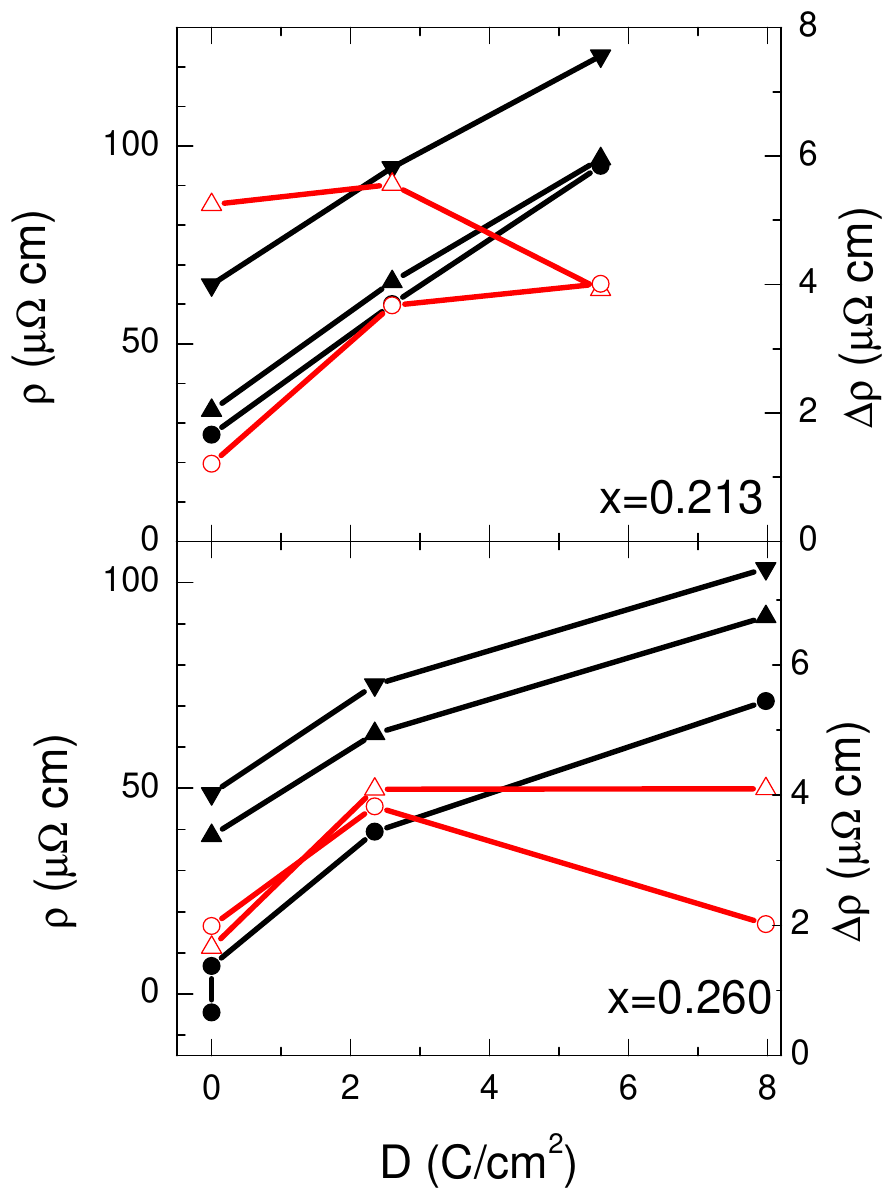}
\caption {(Color online) Irradiation dose dependence of resistivity (left axes, black symbols) and resistivity anisotropy (right axes, red symbols) in samples of Ba$_{1-x}$K$_x$Fe$_2$As$_2$ with $x=$0.213 (top panel) and $x=$0.260 (bottom panel). In the top panel solid black down triangles and open red up-triangles are for $T=$60~K, at about maximum of anisotropy, solid up-triangles and open circles for $T=$22~K, just above $T_c$, and black solid circles in $T\to$0 extrapolation. In the bottom panel solid black down triangles and open red up-triangles are for $T=$55~K, slightly below the $T_{C4}$,  solid black up-triangles and open red circles are for $T=$38~K above  $T_{C4}$, and solid black circles for $T=$0 extrapolation.  }
\label{dosedependence}
\end{figure}

In Fig.~\ref{dosedependence} we show evolution of the resistivity and of the resistivity anisotropy at characteristic temperatures with irradiation dose. For sample with $x=$0.213 these temperatures were selected as $T=$60~K (in the vicinity of the maximum of anisotropy), at $T=$22~K (above onset of the superconducting transition) and in $T\to$0 extrapolation. It is known that resistivity at a fixed temperature in irradiation chamber changes linearly with dose \cite{npjQM,PRX}, the Matthiessen rule is strongly violated in nearby $x=$0.20 composition. Interestingly, resistivity in $T \to$0 extrapolation varies almost perfectly linearly with dose (black solid circles), but downward deviation from linear trend is found at 22~K and 60~K. Resistivity anisotropy at 60~K remains relatively constant. Resistivity anisotropy above $T_c$ initially rises, then seems to saturate.

For sample with $x=$0.260 (bottom panel in Fig.~\ref{dosedependence}) the resistivity increase for all temperatures has a tendency to downward deviation. One possibility is that this is an artefact of incorrect dose determination. Big doses are accumulated over several irradiation runs (during a period up to three years) and partial defect annealing can be happening over these long periods.

\begin{figure}
\includegraphics [width=8.3cm]{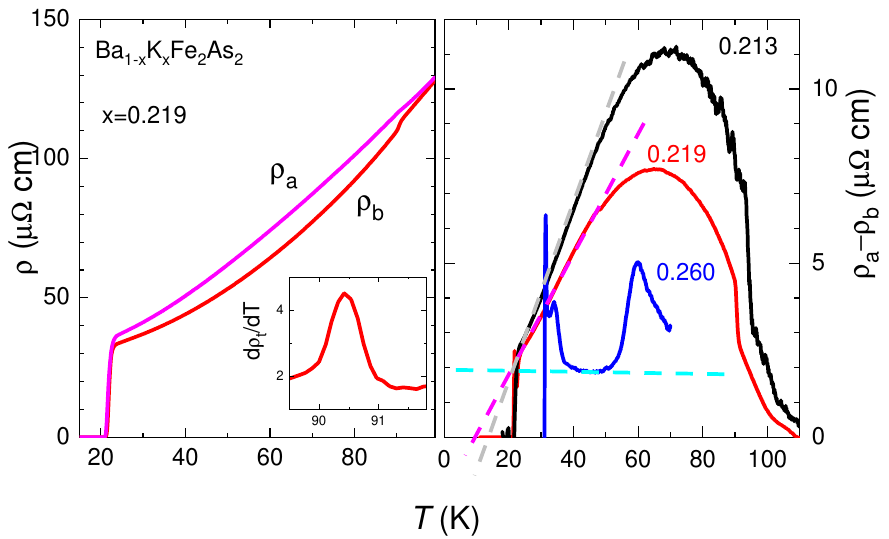}
\caption {(Color online) Temperature-dependent electrical resistivity of Ba$_{1-x}$K$_x$Fe$_2$As$_2$ sample with $x=$0.219 for measurements along $a-$, $\rho_a$ (top curve), and $b-$, $\rho_b$ (bottom curve), directions in the conducting plane (left panel). Inset shows zoom of the structural/magnetic transition. Right panel shows temperature-dependent in-plane resistivity anisotropy, $\Delta \rho \equiv \rho_a-\rho_b$. For reference we show similar measurements in samples $x=$0.213 (black top curve) and $x=$0.260 (bottom blue curve). Dashed lines are guides for eyes.
 }
\label{stressx0p219}
\end{figure}

To check for systematics of the results, we measured one more pristine sample of Ba$_{1-x}$K$_x$Fe$_2$As$_2$ from the same batch as sample $x=$0.213,  however, with somewhat different composition, $x=$0.219. The temperature-dependent electrical resistivity of the stress-detwinned sample with $x=$0.219 for measurements along principal in-plane directions, $\rho_a$ and  $\rho_b$, is shown in Fig.~\ref{stressx0p219}. The sample is characterised by somewhat lower $T_{C2}$ compared to sample $x=$0.213 (inset in left panel of Fig.~\ref{stressx0p219}, 90.6~K vs 94~K) and higher $T_c$, 21.3~K vs 19.8~K.  The resistivity curves show the same tendency as found in pristine sample with $x=$0.213, with two curves converging on cooling above $T_c$. In the right panel of Fig.~\ref{stressx0p219} we show $\Delta \rho(T)$ for sample with $x=$0.219 (red line) in comparison with samples $x=$0.213 (black top curve) and $x=$0.260 (bottom blue curve). We can clearly see two trends with increasing $x$, the decrease of the maximum anisotropy and decrease of the slope of the linear portion of $\Delta \rho (T)$ (highlighted by lines serving as guides for eyes).

\section{Discussion}

There are two main groups of theories explaining nematic resistivity anisotropy, see \cite{Fernandesreview} for the review. The first group is relating the nematic anisotropy to the Drude term, $n/m^*$, reflecting anisotropy of the band structure. The other group of theories is relating $\Delta \rho$ to the anisotropy of scattering, both elastic and inelastic. In all theories the anisotropy should be proportional to the nematic order parameter, $\delta=(a-b)/(a+b)$, as found in scattering \cite{Avci} and thermal expansion measurements \cite{BohmerNC} , the later shown in Fig.~\ref{stressx0p213} and Fig.~\ref{stress0p26} for samples with $x=$0.22 and $x=$0.262, respectively. It is also possible to have a temperature dependent pre-factor $\Upsilon$, coming, for example, from temperature dependent scattering in which case it should be proportional to $\rho$. The analysis of nematic resistivity anisotropy using this approximation, $\Delta \rho=\rho_t \delta$, was very successful in FeSe \cite{FeSe}, giving quite good description of the data. We need to keep in mind though, that the situation in FeSe is simpler than in the hole doped Ba$_{1-x}$K$_x$Fe$_2$As$_2$. Nematic order is not accompanied by the long range magnetic ordering in FeSe, and thus no Fermi surface folding effects are involved \cite{folding1,folding2}. On the contrary, the Fermi surface changes at the transition are important for the hole doped compositions studied here.

\begin{figure}
\includegraphics [width=8.3cm]{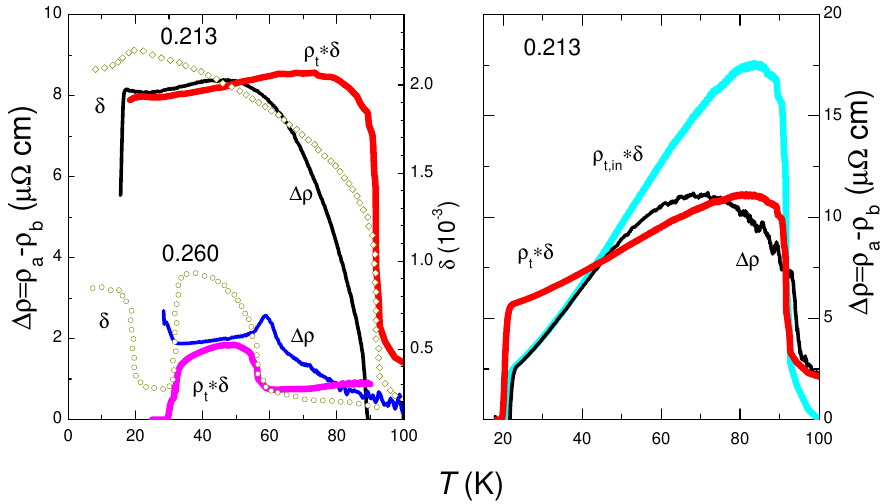}
\caption { (Color online) Left panel. Comparison of the in-plane resistivity anisotropy in samples of Ba$_{1-x}$K$_x$Fe$_2$As$_2$ with $x=$0.213 (5.6 C/cm$^2$) and $x=$0.260 (2.35 C/cm$^2$) with the degree of orthrhombic distortion, $\delta= (a-b)/(a+b)$, as determined in thermal expansion measurements by B\"ohmer {et al.} \cite{BohmerNC} (open yellow circles) and a product $\rho_t \delta$ (red and magenta lines for 0.213 and 0.260, respectively). Right panel. Comparison of $\Delta \rho (T)$ in the pristine sample of $x=$0.213 (black line) with a product $\rho_t \times \delta$ (red line) and of the inelastic part of resistivity, $\rho_{t,in}=\rho_t-\rho_{t,0}$, and the orthorhombic order parameter, $\rho_{t,in}\times \delta$ (cyan line). }
\label{comporderprameter}
\end{figure}

We start with analysis of the heavily irradiated samples, as shown in the left panel of Fig.~\ref{comporderprameter}. Here we compare directly $\Delta \rho (T)$ of the sample with $x=$0.213 irradiated with 5.6 C/cm$^2$ (black line) with $\delta (T)$ measured by B\"ohmer (dark yellow circles) and a product of resistivity in the twinned state $\rho_t(T)$ and  $\delta(T)$   (dark yellow line). For reference we show $\Delta \rho(T)$ for 2.35~C/cm$^2$ irradiated sample with $x=$0.260 (blue line) and $\delta(T)$ for sample with $x=$0.262. First, we can clearly see that the magnitude of the resistivity anisotropy scales with the degree of the orthorhombic distortion $\delta$, in sharp contrast with the electron-doped side \cite{Ian1}. Second, the product $\delta \rho_t$ gives quite good description of the data for $x=$0.22 sample (red vs black curve) below approximately 60~K. The difference at higher temperatures is quite notable, however, it is natural that $\Delta \rho (T)$ has a contribution from temperature dependent folding gap opening.
In the right panel of Fig.~\ref{comporderprameter} we perform the same analysis for sample $x=$0.213 in pristine the state. The resistivity anisotropy $\Delta \rho (T)$ (black line) shows close to $T-$linear dependence. The product $\rho_t\delta$ (we use the same $\delta (T)$ as shown in the left panel) captures this $T-$ linear dependence, despite neither $\rho_t(T)$ (black line in Fig.~\ref{stressfreex0p213}) nor $\delta (T)$ showing $T-$linear dependence. The difference with irradiated case is quite notable, since $\Delta \rho(T)$ decreases notably faster than the $\rho_t\delta$ product in the range where the temperature-dependent folding gap opening should have minor effect.
The match becomes significantly better if we use only inelastic part of the resistivity, $\rho_{t,in}=\rho_t(T)-\rho_t(0)$, as shown with cyan line.

\begin{figure}
\includegraphics [width=8.3cm]{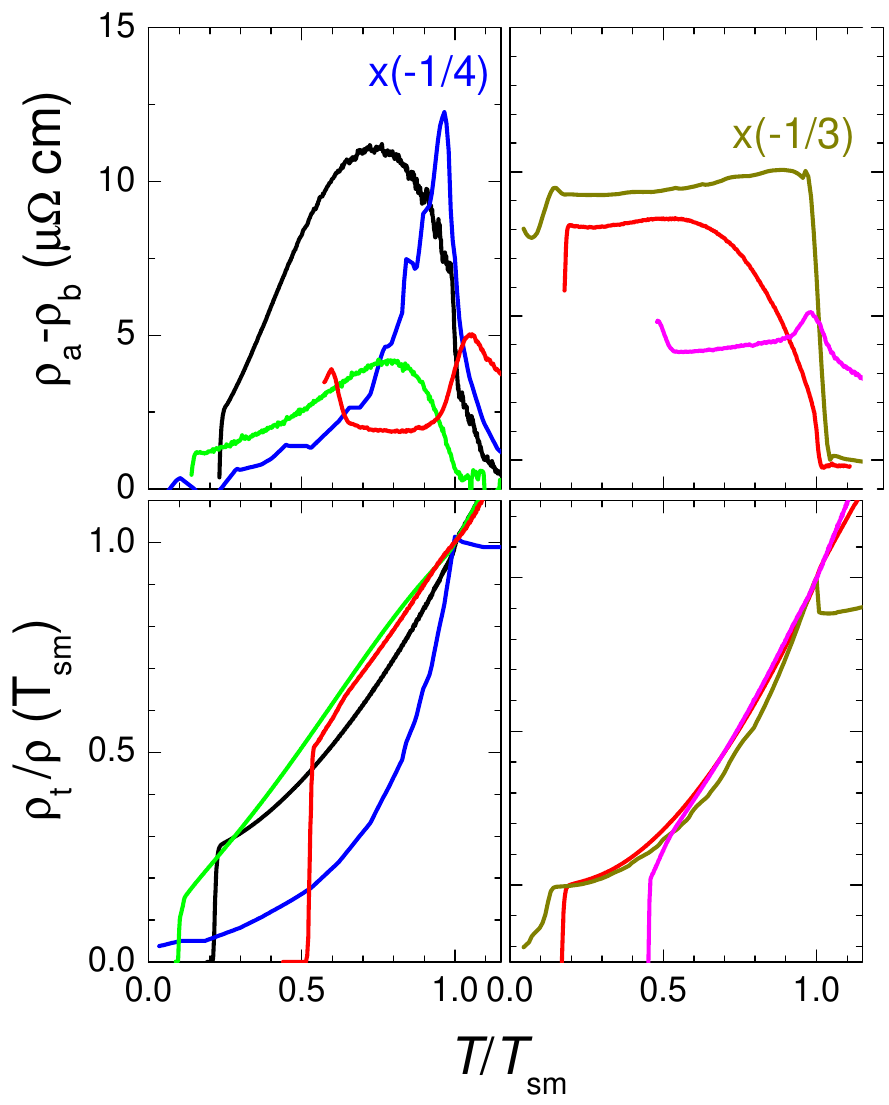}
\caption {(Color online) Comparison of the temperature-dependent electrical resistivity anisotropy for clean samples (left top panel) and dirty (top right) samples of various iron based superconductors. The data are presented vs normalized temperature scale, $T/T_{C2}$. Blue curve is for annealed parent BaFe$_2$As$_2$  \cite{Ishida}, the data are divided by 4, green for FeSe \cite{FeSe}, black and red are for Ba$_{1-x}$K$_x$Fe$_2$As$_2$ samples $x=$0.213 and $x=$0.260, respectively (this study). Yellow curve in the top right panel is for Ru-substituted BaFe$_2$As$_2$ \cite{BlombergJPCM} and is divided by 3. Red line is for the $x=$0.213 irradiated with 5.6 C/cm$^2$, magenta line is for $x=$0.260 sample irradiated with 2.35 C/cm$^2$. The bottom panels show the temperature dependent resistivities of the same compounds, plotted using normalized $\rho(T)/\rho(T_{C2})$ and $T/T_{C2}$ scales.  }
\label{comparisons}
\end{figure}

As a general remark, we should point out, that electron irradiation at the doses used in this study does not introduce variation of carrier density sufficient to have any noticeable impact. This was verified through Hall effect measurements on samples with $x=$0.20 \cite{npjQM} and is in line with common expectations for metals \cite{PdTe}. So for our discussion we can consider effect only through scattering rate.

The results of this study are in general agreement with the previous studies using annealing to control residual resistivity or the samples with naturally low residual resistivity. For example, the decrease of anisotropy from a large value below $T_{C2}$ on cooling to low temperatures is found in perfectly annealed BaFe$_2$As$_2$ \cite{Ishida} (blue curve in Fig.~\ref{comparisons}) and in very clean samples of FeSe \cite{FeSe} (green curve in Fig.~\ref{comparisons}). We explicitly compare the anisotropy found in these compounds with Ba$_{1-x}$K$_x$Fe$_2$As$_2$ samples $x=$0.213 and $x=$0.260 in the pristive state. It was argued \cite{Ishida,BlombergJPCM} that the decreasing anisotropy on cooling is determined by contribution of light carriers \cite{Diraccones1,Diraccones2,Diraccones3,anisotropyDirac1,anisotropyDirac2}, strongly suppressed by disorder scattering. In this respect, close to $T-$ linear dependence of $\Delta \rho (T) $ in the pristine samples with $x=$0.213 and 0.219 may suggest that this group of carriers suffers critical scattering on approaching $C4$ phase boundary. Indeed fluctuations of nematic order parameter with notable contribution of $q=$0 component should have notably bigger effect on small pockets of the Fermi surface.

Strikingly, the increase of residual resistivity with irradiation does not increase anisotropy beyond its maximum value in the clean samples. This fact suggest that $\rho(0)$ does not contribute much to the anisotropy, at least on the hole doped side close to $C_4(x)$ phase boundary.

Interestingly, while $T-$linear dependence is a hallmark of a quantum critical point in the phase diagram of iso-valently substituted BaFe$_2$(As,P)$_2$ \cite{MatsudaP,TanatarP} and partially electron-doped Ba(Fe,$TM$)$_2$As$_2$ \cite{Taillefer}, the temperature-dependent resistivity in Ba$_{1-x}$K$_x$Fe$_2$As$_2$ does not reveal it \cite{YLiucrystals}. Our observation may be suggesting that the reason for this may be phase competition. Indeed, the resistivity in the $C_2$ phase in the sample with $x=$0.260 is close to linear, though in a very limited temperature range.

\section{Conclusions}

The sign reversal of resistivity anisotropy in the samples of hole-doped Ba$_{1-x}$K$_x$Fe$_2$As$_2$ on approaching the reentrant tetragonal phase is insensitive to disorder, opposite to some theory suggestion \cite{inelastic2}. The anisotropy at high temperatures does not depend on the residual resistivity, the anisotropy  of clean samples with $x=$0.213 and 0.219 notably decreases on cooling in the pristine samples and stays constant in the samples with high residual resistivity. This study suggests that inelastic scattering responsible for the temperature-dependent part of resistivity is anisotropic, while elastic scattering responsible for residual resistivity is notably less anisotropic. The temperature dependent anisotropy in pristine samples suggests contribution of high mobility carriers subject to scattering on nematic fluctuations.

\acknowledgments

This research was supported by the U.S. Department of Energy, Office of Basic Energy Sciences, Division of Materials Sciences and Engineering. Ames Laboratory is operated for the U.S. Department of Energy by Iowa State University under Contract No.~DE-AC02-07CH11358.
Irradiation realized on SIRIUS platform was supported by French National network of accelerators for irradiation and analysis of molecules and materials EMIR\&A under project 18-5354.

\end{document}